\documentstyle[prl,aps,epsf]{revtex}
\twocolumn
\tighten

\begin{document}
\draft
\twocolumn[
\hsize\textwidth\columnwidth\hsize\csname@twocolumnfalse\endcsname

\author{R.S. Freitas and L. Ghivelder*}
\address{Instituto de F\'{i}sica, Universidade Federal do Rio de Janeiro, C.P. 68528,
\\
Rio de Janeiro, RJ 21945-970, Brazil}
\author{P. Levy and F. Parisi}
\address{Departamento de F\'{i}sica, Comision Nacional de Energia Atomica, Av. Gral\\
Paz 1499 (1650) San Martin, Buenos Aires, Argentina}
\title{Magnetization studies of phase separation in La$_{0.5}$Ca$_{0.5}$MnO$_{3}$}
\maketitle

\begin{abstract}
We present magnetization studies in a series of phase separated La$_{0.5}$Ca$%
_{0.5}$MnO$_{3}$ manganite samples, with different low temperature fractions
of the ferromagnetic (FM) and charge ordered-antiferromagnetic (CO-AFM)
phases. A particular experimental procedure probes the effect of the
magnetic field applied while cooling the samples, which promotes FM fraction
enlargement and enhances the melting of the CO phase. The response of the
system depending on its magnetic field history indicates the existence of
three different regimes in the phase separated state which develops below T$%
_{C}$. Our data allows us to identify the onset temperature below which the
system becomes magnetic and structurally phase separated, and an onset field
above which FM fraction enlargement occurs.
\end{abstract}
]
\narrowtext

\pacs{75.30.Vn, 72.80.Ga, 75.30 Kz}

\section{Introduction}

The mixed-valent manganites R$_{1-x}$A$_{x}$MnO$_{3}$ (R = rare-earth, A =
Ca, Sr, or Ba) attracted considerable scientific interest due to their wide
variety of spin, charge, and orbital states.\cite{Salamon} Extensive
investigation in these compounds was first stimulated by the discovery of
colossal magnetoresistance (CMR), a large decrease in electrical transport
when a magnetic field is applied, which takes place in various compounds
near a ferromagnetic transition. Subsequently, the main focus of attention
has moved beyond the study of CMR effects in manganites, in particular to
the {\it x} = $\frac{1}{2}$ substituted compositions, which in La$_{1-x}$Ca$%
_{x}$MnO$_{3}$ represents the boundary between competing ferromagnetic (FM)
and charge ordered-antiferromagnetic (CO-AFM) ground states, a favorable
scenario for phase separation (PS) phenomena.\cite{Moreo} Phase separation
also occurs in several other systems, and it is most studied in (La$_{1-z}$Pr%
$_{z}$)$_{0.7}$Ca$_{0.3}$MnO$_{3}$ at an intermediate {\it z} doping range;%
\cite{Uehara} the end members {\it z} = 0 and {\it z} = 1 are FM metallic
and CO-AFM insulator at low temperatures, respectively. Recent experimental
investigations have shown undisputed evidence of the existence of
inhomogeneous states and PS in various manganite compounds,\cite{Dagotto} a
major discovery in the study of strongly correlated electron systems.

In the case of La$_{0.5}$Ca$_{0.5}$MnO$_{3}$, the ``standard'' compound
changes on cooling to a FM phase at T$_{C}$ $\sim $ 220 K, and subsequently
to CO-AFM at T$_{CO}$ $\sim $ 150 K ($\sim $ 180 K upon warming).\cite
{Schiffer} However, it has been established that this system is better
described as magnetically phase-segregated over a wide range of
temperatures. The nature and properties of this phase separated state have
been the subject of numerous experimental\cite
{Huang,Levy,Mori,Allodi,Yoshinari,Parisi} and theoretical\cite{Moreo,Dagotto}
investigations. At low temperatures, T 
\mbox{$<$}%
T$_{CO}$, FM metallic regions are trapped in a CO-AFM matrix. This is
consistent with the sizable bulk magnetization values observed,\cite
{Huang,Levy} arising from the ferromagnetic component, and with resistivity
data, which may show metallic behavior due to percolation of the FM
clusters. The fraction of FM phase remaining at low temperatures is strongly
sample dependent. It has been previously demonstrated\cite{Levy} that the
fraction of FM and CO-AFM phases in polycrystalline La$_{0.5}$Ca$_{0.5}$MnO$%
_{3}$ can be controlled by appropriate thermal treatments, which produces a
continuous change of the relative fraction of the competing FM and CO-AFM
phases. The existence of phase separation within the FM phase, i.e., in the
intermediate temperature region T$_{CO}$ 
\mbox{$<$}%
T 
\mbox{$<$}%
T$_{C}$, has also been established in various investigations. The occurrence
of an incommensurate charge/orbital ordered state between T$_{C}$ and T$%
_{CO} $, incompatible with ferromagnetic order, was found through
transmission electron microscopy.\cite{Mori,Chen-Cheong} Detailed neutron
studies\cite{Huang} reported the appearance below T$_{C}$ of a second
crystallographic phase structurally different from the FM phase, and lacking
magnetic order. A consensus has now emerged in the literature regarding the
paramagnetic character of this secondary phase,\cite
{Simopoulos,Rivadulla,Mahendiran} although the possible existence of an AFM
incommensurate charge-ordered state can not be ruled out.\cite{Mori}

Microscopic probes were more successful than magnetic measurements in
detecting the intermediate phase separated state. Indeed, previous
magnetization studies appeared to be consistent with the existence of a
homogeneous ferromagnetic state at T$_{CO}$ 
\mbox{$<$}%
T 
\mbox{$<$}%
T$_{C}$ ,\cite{Schiffer,Levy,Roy} without any indication of PS. The main
reason for the mentioned limitation of the magnetic techniques can be found
in the sensitivity of the phase separated state to the measuring magnetic
field,\cite{Allodi,Yoshinari,Ritter} which can mask the real characteristics
of the system in conventional magnetization measurements. In a previous
investigation\cite{Parisi} we have shown through transport measurements how
the application of a low magnetic field can alter the volume fraction of the
competing magnetic phases. When the field is applied while cooling the
sample, in a conventional field-cooled-cooling (FCC) experiment, it promotes
an enlargement of the FM volume fraction, by preventing the formation of
otherwise non-FM regions. On the other hand, if the field is turned on while
the measurement is made and turned off while cooling the sample, a procedure
called {\it turn-on-turn-off }mode, the relative fraction of the existing
phases is less disturbed during the cooling process. The measured
magnetoresistance of the compounds differ considerably when comparing data
obtained with the FCC and {\it turn-on-turn-off} modes, since the former
enhances the percolation of the FM metallic regions. In fact, it has been
theoretically proposed\cite{Dagotto,DagottoPRL} and experimentally confirmed%
\cite{Parisi,Mahendiran} that the observed colossal magnetoresistance effect
in manganites is caused by a field-induced percolative transition of
metallic regions in phase separated systems. Within this scenario, it is
possible that a moderate applied field ($\sim $ 1 T) can be high enough to
suppress the non-FM phase between T$_{C}$ and T$_{CO}$ in a FCC experiment,
giving rise to a magnetic response of the system consistent with a
homogeneous state. This is probably the reason for the lack of detection of
the intermediate phase separated state through magnetic measurements.

In this paper we present a detailed magnetization study of phase separated La%
$_{0.5}$Ca$_{0.5}$MnO$_{3}$ manganites, focusing mainly in the intermediate
temperature region T$_{CO}$ 
\mbox{$<$}%
T 
\mbox{$<$}%
T$_{C}$. The samples are a series of polycrystalline compounds, with varying
low-temperature volume fraction of FM and CO-AFM phases. We investigated the
magnetic response of the system under different measurement procedures,
extending the idea related with {\it turn-on-turn-off} -type measurements to
magnetization studies. We found that by changing separately the measuring
and cooling field leads to different magnetic responses, providing an
important tool to investigate PS effects through magnetization measurements.
Using this particular experimental procedure we were able to identify three
different regimes in the phase separated state of La$_{0.5}$Ca$_{0.5}$MnO$%
_{3}$ below T$_{C}$, related to the response of the system to the cooling
and measuring fields.

\section{Experimental Details}

Polycrystalline La$_{0.5}$Ca$_{0.5}$MnO$_{3}$ was obtained by a standard
citrate/nitrate decomposition method. Additional thermal treatments were
performed to obtain a batch of samples with different grain sizes. Care was
taken to ensure that all samples have the same oxygen stoichiometry, and
same Ca doping concentration near 0.5. More details on material preparation
and x-ray analysis were given elsewhere.\cite{Levy} Magnetization
measurements were performed with a commercial magnetometer (Quantum Design
PPMS) between room temperature and 2 K, with applied fields up to {\it H} =
9 T. Transport data was obtained with a standard four-point technique. Table
I lists some physical parameters of the measured samples, labeled A to E:
grain sizes, low temperature volume fraction of the FM phase, ferromagnetic
transition temperature T$_{C}$, and charge-ordered antiferromagnetic
transition temperature T$_{CO}$ measured on cooling. The fraction of FM
phase was estimated from {\it M} vs. {\it H} measurements at 10 K, as
previously described.\cite{Levy}

\section{Results}

In order to characterize the magnetic behavior of the studied compounds,
Fig. 1 shows the temperature dependence of the magnetization, measured with 
{\it H} = 2 mT for samples A to E. As the temperature is lowered all samples
display a FM transition at T$_{C}$, followed by a hysteretic first order
transition to a CO-AFM state at T$_{CO}$ (the latter not visible for sample
A). The bulk magnetization measured below T$_{CO}$ signals the presence of
FM domains trapped in the CO-AFM matrix, as previously reported for this
material.\cite{Huang,Levy} It is clear from the data that the FM volume
fraction strongly increases through the series, starting from a compound
mostly CO-AFM at low temperatures in sample E, and reaching a nearly fully
FM material in sample A. The transition temperatures T$_{C}$ and T$_{CO}$,
obtained through the maximum slope of the magnetization data, are listed in
Table I. The FM transition decreases from T$_{C}$ $\sim $ 227 to 214 K for
samples A to E respectively, and the temperature of the CO-AFM transition
increases from T$_{CO}$ $\sim $ 128 to 142 K, signaling the stabilization of
the CO-AFM phase as the low temperature FM fraction decreases in the
materials. A pronounced irreversibility, i.e., the difference between the
field-cooled-cooling (FCC) and zero field-cooled (ZFC) curves, is observed
for all samples, as a consequence of the magnetic anisotropy of the
compounds. The fractional change $\xi $ in magnetization, expressed as 

\begin{center}
\begin{table}[tbp]
\caption{Main characteristics of La$_{0.5}$Ca$_{0.5}$MnO$_{3}$ samples,
labelled A to E. FM (\%) is the ferromagnetic phase fraction at low
temperatures,  T$_{C}$ the ferromagnetic transition temperature,
and T$_{CO}$ the charge-ordered antiferromagnetic transition temperature
measured on cooling. T$_{C}$ and T$_{CO}$ were obtained from the maximum
inflection of the low-field (2 mT) magnetization data.}
\begin{tabular}{|c|c|c|c|c|} 
Sample & grain size ($\mu $m) & FM (\%) & T$_{C}$ (K) & T$_{CO}$ (K) \\ 
A & 180 & 84 & 225 & $-$ \\ 
B & 250 & 77 & 219 & 129 \\ 
C & 450 & 54 & 218 & 137 \\ 
D & 950 & 15 & 215 & 143 \\ 
E & 1300 & 9 & 214 & 147
\end{tabular}
\end{table}
\end{center} 
 $\xi =(M_{FCC}-M_{ZFC})/M_{FC}$, has approximately the same value in all
samples, $\xi $ $\simeq $ 0.94 $\pm $ 0.03 at 10 K. Magnetization
measurements at various fields (not shown) reveal little changes in $\xi $
up to {\it H} = 0.02 T, and a strong decrease above {\it H} $\simeq $ 0.1 T.
This is consistent with the values obtained for the coercive field (a
measure of the magnetic anisotropy), {\it H}$_{c}$ $\simeq $ 0.05 T at 10 K
for all samples. For {\it H} $\gg $ {\it H}$_{c}$, $\xi $ becomes negligible.

The effect of the magnetic field on the coexisting FM and CO-AFM phases can
be readily visualized through measurements of {\it M} vs. {\it H}. Figure 2
shows data taken at 130 K, just below the CO transition. It is interesting
to observe how the results evolve through the series of samples. For samples
D and E, with higher CO-AFM content at low temperatures, the effects due to
the FM and AFM phases are clearly separated in the curves. As the field
increases it initially aligns the FM moments, followed by a metamagnetic
transition of the AFM moments to a field-induced ferromagnetic phase. At
this temperature, and at the highest field reached, the entire sample is in
a FM state. The metamagnetic transition displays large field hysteresis and
time relaxation at the fields where dM/dH is maximum. Samples C and D show a
larger initial saturation value of the FM moments and less pronounced
hysteresis, related with the increase of the FM fraction and decrease of the
CO-AFM fraction in the compounds. Sample A shows data similar to a fully FM
sample, consistent with the results of magnetization as a function of
temperature (Fig. 1).

\begin{center}
\begin{figure}[tbp]
\hspace{-0.0cm}
\epsfxsize=5.5cm
\centerline{\epsffile{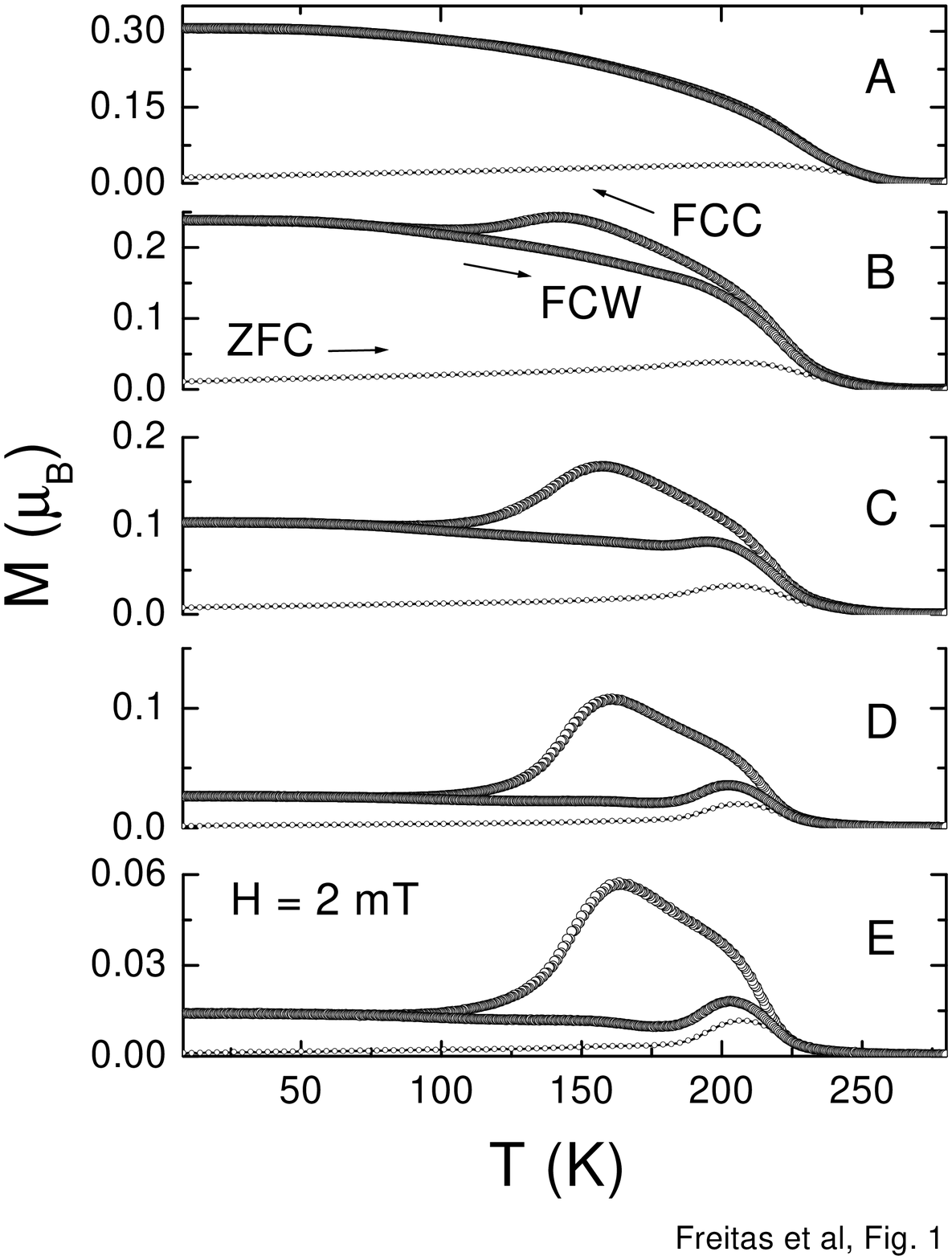}}
\caption{Magnetization as a function of temperature of La$_{0.5}$Ca$_{0.5}$%
MnO$_{3}$ samples, labelled A to E. Results obtained with {\it H} = 2 mT,
with zero-field-cooling (ZFC), field-cooled-cooling (FCC), and
field-cooled-warming (FCW) procedures.
\label{Fig1} 
}
\end{figure}
\end{center}

\begin{center}
\begin{figure}
\hspace{-0.0cm}
\epsfxsize=5.50cm
\centerline{\epsffile{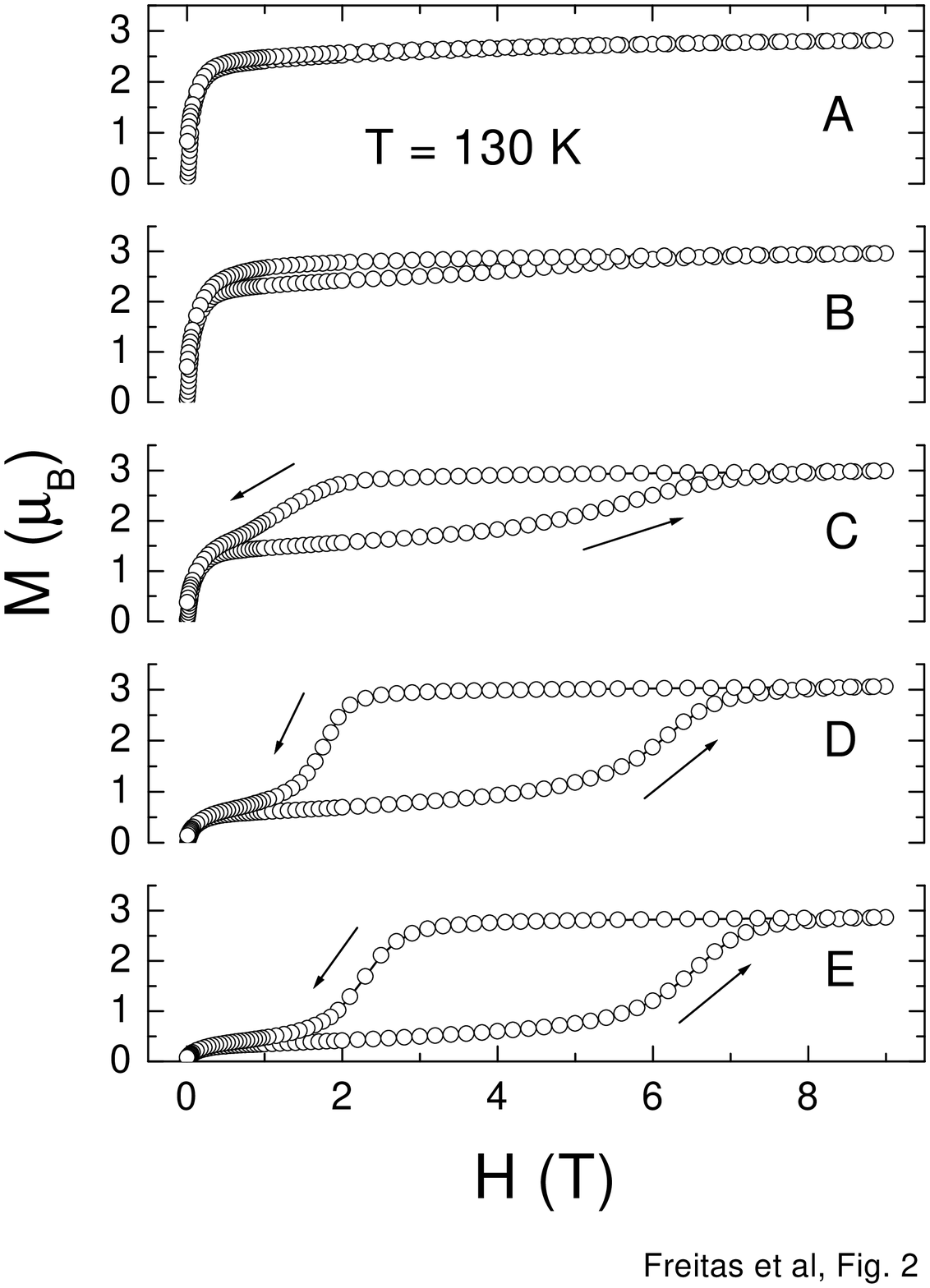}}
\caption{Magnetization as a function of field of La$_{0.5}$Ca$_{0.5}$MnO$%
_{3}$ samples, labelled A to E. Results obtained after zero-field-cooling to
T = 130 K.
\label{Fig2} 
}
\end{figure}
\end{center}

We shall now demonstrate that when a magnetic field is applied while cooling
the samples it can dynamically change the relative fraction of the magnetic
states in this phase separated material. This phenomena is investigated by
extending to magnetization data the idea of {\it turn-on-turn-off} -type
measurements, where the field is left on at fixed temperatures while taking
a data point, and turned off while cooling the sample. In a more general
procedure employed in the present study, the measured magnetization may
depend on two different variables, {\it x} and {\it y}, where {\it x} is the
measuring field and {\it y} the field in which the sample is cooled (both
expressed in T). This measuring technique is hereafter called the {\it M}(%
{\it x-y}) mode. Figure 3 displays the results obtained for all samples,
comparing the data taken with the FCC and {\it M}(1-0) modes, the latter
with a measuring field of 1 T and cooling field zero. The value of 1 T was
chosen since it is much above the observed anisotropy field and much below
values that would cause a pronounced decrease in T$_{CO}$ due to melting of
the CO state. For sample A, nearly fully FM down to low temperatures, the
FCC and {\it M}(1-0) results virtually coincide. As the CO-AFM fraction
increases through the series the magnetization becomes strongly dependent on
whether or not the field is left on while cooling the samples. In samples B
and C, below a certain temperature, the {\it M}(1-0) data is shifted with
respect to the FCC data. In samples D and E, with higher CO-AFM fraction, it
is remarkable to observe huge differences in bulk magnetization when
comparing the curves obtained with the two measuring procedures. The
temperature where the FCC and {\it M}(1-0) curves separate, T$_{o}$ $\simeq $
200 $\pm $ 2 K, is nearly unchanged in these two samples,
\twocolumn[
\hsize\textwidth\columnwidth\hsize\csname@twocolumnfalse\endcsname
\begin{center}
\begin{figure}
\hspace{-0.0cm}
\epsfxsize=12.50cm
\centerline{\epsffile{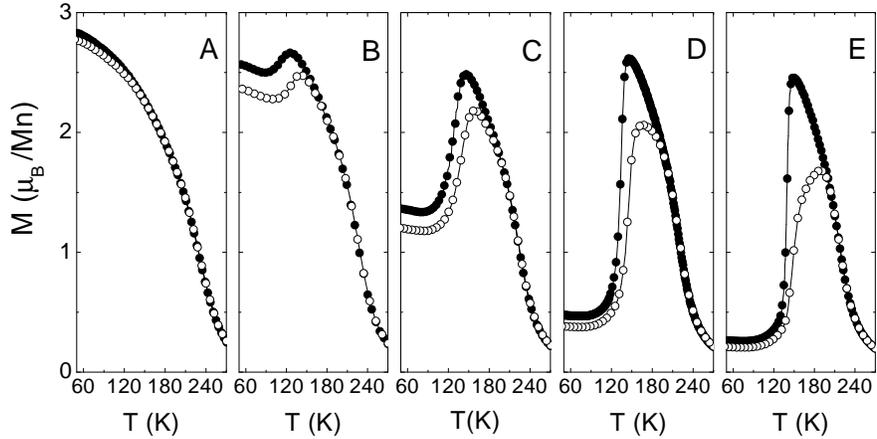}}
\caption{Magnetization as a function of temperature of La$_{0.5}$Ca$_{0.5}$%
MnO$_{3}$ samples, labelled A to E. The curves with solid symbols were
measured with a field-cooled-cooling (FCC) mode, with H = 1T. Curves with
open symbols obtained with a measuring field {\it x} = 1 T, and a cooling
field {\it y} = 0, called the {\it M}(1-0) mode.
\label{Fig3} 
}
\end{figure}
\end{center}
]
 and progressively
decreases through the series. As discussed below, the change in
magnetization depending on the measuring mode is due to the existence of
non-FM regions within the FM phase.

Before interpreting these results some points must be verified to ensure the
physical meaning of the data. Since the field is being switched on and off
continuously in the {\it M}({\it x-y}) mode, one may suspect the samples to
be in some kind of magnetic metastable state, where strong relaxation
effects would be present. In order to rule out this possibility the {\it M}%
(1-0) data was repeated leaving the field on for 30 minutes, and
subsequently leaving the field switched off for the same period of time
before proceeding to next data point. The observed fractional change in {\it %
M}(1-0) is always less than 2 \% at all temperatures. Long time relaxation
data was also taken at selected temperatures, giving a fractional change in 
{\it M}(1-0) of less then 3 \% in a period of 4 hours. A different test was
also made, by varying the temperature interval between data points, from 0.5
to 5 K, with no effect in the {\it M}(1-0) curves. These results clearly
establish that the field in which the samples are cooled has an overwhelming
effect in the bulk magnetization measured. It is worth noting that the
results (not shown) taken while warming the samples, with
field-cooled-warming and {\it M}(1-0) warming modes, virtually coincide.

The remaining of this paper will focus on measurements on sample E, with the
largest CO-AFM volume fraction at low temperatures and where PS effects are
more pronounced. Figure 4 shows additional measurements taken in the M({\it %
x-y}) mode. The main panel displays the results at intermediate
temperatures, while the inset shows an enlarged plot with the low
temperature data. The measuring field was kept constant at x = 1 T, and the
cooling field {\it y} was varied in order to study its influence on the
measured magnetization. The {\it M}(1-0.06) data (not shown for clarity of
the figure) is virtually identical to {\it M}(1-0); a small difference
between {\it M}(1-0.1) and {\it M}(1-0) is observed, and the {\it M}({\it x-y%
}) curves start to clearly separate from the {\it M}(1-0) data at {\it y} $%
\gtrsim $ 0.2 T. With increasing cooling fields {\it y} the curves gradually
evolve towards the data obtained with the FCC mode. With the measuring field
kept constant, these results indicate an onset at {\it H} $\simeq $ 0.1 T
above which the cooling field induces changes in the measured magnetization.
This is consistent with the results shown in Fig. 5a, where it is shown that
with a measuring field of 0.1 T, FCC and {\it M}(0.1-0) results practically
coincide. The different panels in Fig. 5 compare data taken with the FCC and 
{\it M}({\it x}-0) modes. The cooling fields is kept at zero, and the
measuring field {\it x} increases from 0.1 to 0.3, 1 and 3 T respectively.
As will be discussed below, the data in the each of the panels of Fig. 5 may
be interpreted in a qualitatively different way. The same overall results
were obtained on samples B to D, which shows that the cooling field effect
is the same regardless of the ratio between the competing FM and CO-AFM
phases.

\begin{center}
\begin{figure}[tbp]
\hspace{-0.0cm}
\epsfxsize=7.0cm
\centerline{\epsffile{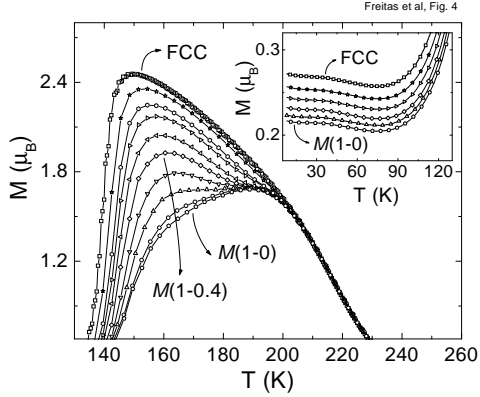}}
\caption{Magnetization as a function of temperature of La$_{0.5}$Ca$_{0.5}$%
MnO$_{3}$, sample E. Results obtained with a measuring field {\it x} = 1 T
and several values of the cooling field {\it y}, the {\it M}(1-{\it y})
mode. The main panel shows an enlarged portion of the data at intermediate
temperatures, with (from bottom to top) {\it y} = 0, 0.1, 0.2, 0.3, 0.4,
0.5, 0.6, 0.7, and 0.85 T. The inset shows the same results at lower
temperatures, with {\it y} = 0, 0.2, 0.4, 0.6, and 0.85 T.
\label{Fig4} 
}
\end{figure}
\end{center}

\begin{center}
\begin{figure}[b]
\hspace{-0.0cm}
\epsfxsize=8.0cm
\centerline{\epsffile{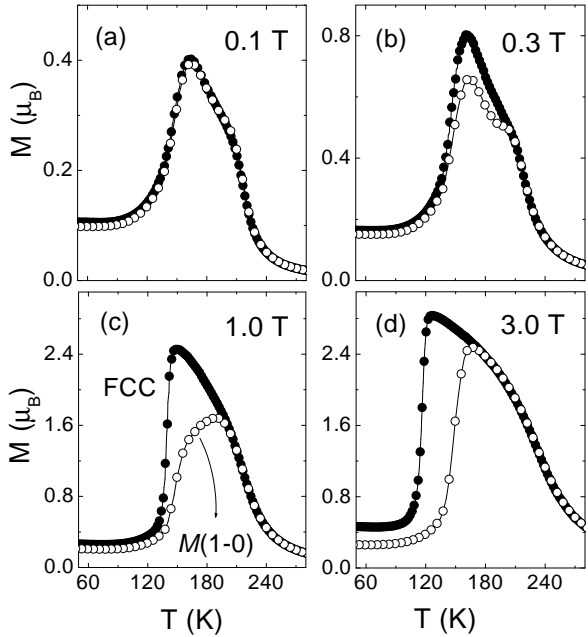}}
\caption{Magnetization as a function of temperature of La$_{0.5}$Ca$_{0.5}$%
MnO$_{3}$, sample E. The curves with solid symbols were measured with a
field-cooled-cooling (FCC) mode, with H = 0.1T (a), 0.3 T (b), 1 T (c) and 3
T (d). Curves with open symbols measured with a cooling field {\it y} = 0,
the {\it M}({\it x}-0) mode.
\label{Fig5} 
}
\end{figure}
\end{center}

Complementary resistivity measurements are shown in Figure 6. Three
different curves are plotted, corresponding to different experimental
procedures: measured without an applied field ({\it H} = 0), measured with 
{\it H} = 1 T on FCC mode, and measured with a field {\it x} = 1 T while
cooled in zero field, the {\it R}(1-0) mode. The inset of Fig. 6 depicts the
magnetoresistance, $MR=[{\it R}(0)-{\it R}({\it H})]/{\it R}(0)$, comparing
the results obtained with FCC and {\it R}(1-0) modes. The results clearly
show that the large low temperature {\it MR} is solely due to the cooling
field, as discussed in more detail in the following section.

\begin{figure}[tbp] 
\begin{center}
\hspace{-0.0cm}
\epsfxsize=7.50cm
\centerline{\epsffile{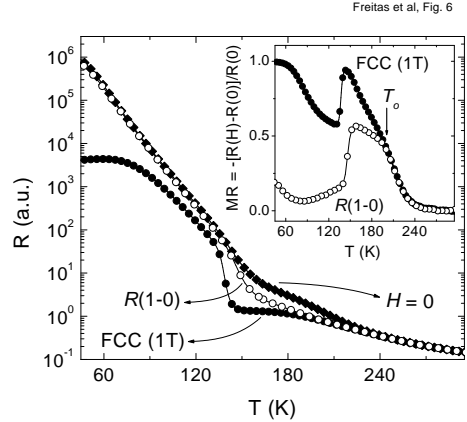}}
\caption{Electrical resistance of La$_{0.5}$Ca$_{0.5}$MnO$_{3}$, sample E.
Results measured with {\it H} = 0, field-cooled-cooling with {\it H} = 1 T,
and with a measuring field {\it x} = 1 T and cooling field {\it y} = 0, the 
{\it R}(1-0) mode. The inset depicts the magnetoresistance $MR=-[{\it R}(H)-%
{\it R}(0)]/{\it R}(0)$, calculated for the FCC and {\it R}(1-0) modes; the
curves separate at T$_{o}$.
\label{Fig6} 
}
\end{center}  
\end{figure}

\section{Discussion}

When performing magnetic measurements in phase separated materials the
applied magnetic field can affect the balance between the coexisting phases
by increasing the amount of the FM phase, an effect called ferromagnetic
fraction enlargement (FFE).\cite{Parisi,Mahendiran,Sacanell} The
magnetization response of a phase separated system involves two field
dependent contributions: the alignment of the magnetic moments of the FM
phase and the FFE effect. The applied magnetic field can also lower T$_{CO}$%
, a phenomena known as melting. It has been argued that PS plays a major
role in the melting mechanism, with the competition and coexistence among
the FM component and the CO phase causing a drop in the CO transition
temperature.\cite{Younghun} Our results (Figs. 3 to 6) clearly show that the
capability of the magnetic field to change the relative fractions of the
coexisting phases is strongly dependent on the way in which the field is
applied, with the cooling and measuring fields affecting the system in
different ways. As previously reported,\cite{Parisi} in a FCC procedure the
cooling field induces FFE mainly by inhibiting the formation of the non-FM
regions. On the other hand, the measuring field, applied solely at fixed
temperatures once the zero-field relative fractions are established, can
produce FFE only if it is large enough to transform some fraction of the
material from one phase to the other, in a metamagnetic-like transition. It
must be emphasized that the observed difference in the magnetization curves
regarding the way in which the field is applied is a direct consequence of
the phase separated nature of the system. On a nearly homogenous FM sample
the magnetization is almost the same in the {\it M}({\it x-y}) and FCC
modes, as observed in Fig. 3(a).

The results of Fig. 3 emphasize the fact that regardless of the relative
low-temperature fractions of the FM and CO-AFM phases, which is highly
dependent on sample preparation, there is an onset temperature T$_{o}$ below
which the cooling field plays an unbalancing role in favor of the FM state.
Measurements displayed in Fig. 4 evidence that above a threshold value of 
{\it H} $\simeq $ 0.1 T the FM phase fraction monotonously increases as a
function of the applied cooling field. The same threshold field is obtained
for the other samples (data not shown). These findings are most likely an
intrinsic characteristic of the La$_{0.5}$Ca$_{0.5}$MnO$_{3}$ system, beyond
the particular differences among the samples.

In order to adequately separate the effects of the measuring and cooling
fields in the ground state of the system let us call $\Delta {\it f}(${\it %
x-y}) the FFE produced when measuring in the {\it M}({\it x-y}) mode, and $%
\Delta {\it f}(${\it x}) the FFE induced in a FCC procedure. The actual FM
phase fraction at a given temperature is proportional to $(f_{0}+\Delta f)$,
where $f_{0}$ is the equilibrium fraction of the FM phase at zero field.
This notation is useful for a qualitative analysis of Fig. 5, which provides
interesting insights in to how the magnitude of the applied magnetic field
affects the phase separated state below T$_{C}$ by comparing the results
measured in the {\it M}({\it x}-0) and FCC modes. Above T$_{o}$ the
magnetization response is the same in both modes, up to the highest field
employed (3 T). The equilibrium state of the sample under the applied
magnetic field is the same irrespective of the way in which the magnetic
field is applied, so we can conclude that $\Delta {\it f}(${\it x}-0) $%
\approx $ $\Delta {\it f}(${\it x}) for T 
\mbox{$>$}%
T$_{o}$.

On cooling below T$_{o}$ a different magnetization response arises depending
on the measuring mode and on the applied field. While the curves obtained
with the extreme fields used (0.1 T and 3 T) show that $\Delta {\it f}(${\it %
x}-0) $\approx $ $\Delta {\it f}(${\it x}), a net distinction between FCC
and {\it M}({\it x}-0) modes is observed for 0.3 T and 1 T, indicating that $%
\Delta {\it f}(${\it x}) 
\mbox{$>$}%
$\Delta {\it f}(${\it x}-0) for these intermediate fields. In the data
measured with {\it H} = 0.3 T, shown in Fig. 5b, the difference below T$_{o}$
between the FCC and {\it M}(0.3-0) results give direct evidence of the FFE
induced by the cooling field. While the magnetization in the FCC mode
continuously increases on cooling across T$_{o}$, the {\it M}(0.3-0) curve
shows that a sudden loss of FM phase happens in a small temperature region
close below T$_{o}$. On further cooling {\it M}(0.3-0) increases again,
indicating that the remaining amount of the FM phase is almost constant
until T$_{CO}$. Some remarkable differences are found in the {\it M}(1-0)
curve (Fig. 5c) with respect to the {\it M}(0.3-0). The former also starts
to deviates from the FCC curve at T$_{o}$, but continuously decreases on
further cooling, contrarily to what happens with {\it M}(0.3-0). This fact
can be attributed to the effect of the measuring field, which in this case
is large enough to convert non-FM regions to a FM state. This process is
accomplished each time the field is turned on. As the temperature is lowered
the capability of the field to produce such transformation decreases. Thus,
while $\Delta {\it f}($1T) remains almost constant in the intermediate
temperature range, $\Delta {\it f}($1-0) decreases as T$_{CO}$ is approached
from above. The capability of even higher measuring magnetic fields to
convert the material from non-FM to FM is clearly visualized in Fig. 5d,
where results obtained with the {\it M}(3-0) and FCC modes coincide above T$%
_{CO}$, indicating that the measuring field is sufficient to induce a nearly
complete metamagnetic transition. The above described scenario is consistent
with that proposed by Mahendiran et al.,\cite{Mahendiran} who suggested that
the phase coexistence above T$_{CO}$ is characterized by the existence of CO
domains of different sizes, and concomitantly, different critical field
values for the local metamagnetic transition.

At T$_{CO}$ the system changes to a CE-type CO-AFM state.\cite{Goodenough}
It is well known that below this temperature the phase separated state is
characterized by the coexistence of small FM regions embedded in a CO-AFM
matrix. There is negligible FFE induced by the measuring field, whereas the
cooling field considerably enhances the low-temperature FM fraction (see
inset of Fig 4). Fast cooling rates also produce the same effect.\cite
{Uehara-eulett} It is worth noticing that the charge order transition
temperature T$_{CO}$ is lower on the FCC mode as compared to the {\it M}(%
{\it x}-0) mode, because the cooling field enhances the melting of the
CO-AFM state. The value of T$_{CO}$ (obtained through the maximum slope at
the decrease of the magnetization) ranges from 146 K at {\it H}= 0.1 T to
116 K at {\it H}= 3 T in the FCC mode, but it is almost field independent (T$%
_{CO}$ = 146 $\pm $ 1 K) in the {\it M}({\it x}-0) mode. The physical
processes occurring in each of the panels of Fig. 5 may be summarized in the
following way: 5a) cooling and measuring fields induce no changes in the
magnetic phases; 5b) cooling field induces FM fraction enlargement; 5c)
Cooling field induces FM fraction enlargement and enhanced melting of the CO
transition, while the measuring field induces only partial FFE; 5d) FM
fraction enlargement is saturated by both the cooling and the measuring
fields above T$_{CO}$, cooling field also promotes melting of the CO
transition.

The resistivity results of Fig. 6 are consistent with this analysis. The
onset temperature T$_{o}$ below which FCC and {\it turn-on-turn-off} -type
measurements separate is the same in magnetization and resistivity data.
Above T$_{o}$, FCC and {\it R}(1-0) resistivity results also coincide.
Between T$_{o}$ and T$_{CO}$ the resistivity is lower when cooling in the
presence of a magnetic field. It is remarkable to note that below T$_{CO}$
the magnetoresistance is entirely due to the cooling field. Results with 
{\it H} = 0 and with {\it H} = 1 T applied solely during the measurements
practically coincide below T$_{CO}$. On the other hand, {\it H} = 1 T
applied while cooling the sample considerably lowers the resistivity. The
inset of Fig. 6 shows how the magnetoresistance is enhanced in the FCC mode
as compared to the {\it R}(1-0) mode. This confirms that PS effects are
responsible for the observed magnetoresistance in this compound. Transport
measurements evidence the different ways in which the system reaches the low
temperature CO-AFM state depending on the measuring mode. As the temperature
is lowered in the FCC mode a sharp increase of resistivity is observed at T$%
_{CO}$, indicating that CO and AFM states occur simultaneously when cooling
under a 1 T field. On the other hand, the zero field resistivity curve
changes nearly smoothly while cooling, suggesting that the fully CO state is
reached by a continuous increase of the CO regions below T$_{C}$. The
enhanced melting (lower T$_{CO}$) as a result of the cooling field is also
observed in the magnetoresistance plot.

The overall behavior observed in La$_{0.5}$Ca$_{0.5}$MnO$_{3}$ through
magnetization measurements, and confirmed by resistivity, shows the
existence of well differentiated regimes in the phase separated state below T%
$_{C}$. The main characteristics of each regime are clearly visualized in
the {\it M} vs. {\it H} data showed in Fig. 7. At different temperatures we
performed the measurements with a definite experimental procedure: the
sample is cooled without an applied field to the target temperature, or
alternatively, the sample is cooled with {\it H }= 1 T to the target
temperature. At 210 K (top inset of Fig. 7), in the high temperature region T%
$_{o}$ 
\mbox{$<$}%
T 
\mbox{$<$}%
T$_{C}$, both results coincide. The inhomogeneous FM state that develops at T%
$_{C}$ is characterized by the coexistence of isolated FM clusters embedded
in a paramagnetic host.\cite{Simopoulos,Rivadulla} With an applied field
these clusters can grow freely against the paramagnetic phase,
irrespectively of the way in which the field is applied.

The temperature T$_{o}$ signals the onset of a different phase separated
state. Following the results of Huang et al.,\cite{Huang} this phase
separated state consists of FM regions coexisting with a structurally
different phase. The second phase has different cell parameters than the FM
one, and is characterized by a major degree of order of the d($_{z}$ $^{2}$)
orbital. Traces of PS of a structural nature, with an onset at T $\approx $
200 K, have also been found in magnetostriction studies.\cite{Mahendiran-SSC}
Thus, below T$_{o}$, the FM clusters are structurally confined and can not
grow freely when a magnetic field is applied in the {\it M}({\it x-y}) mode,
due to the presence of energy barriers at the phase boundaries. Magnetic
relaxation measurements in the intermediate temperature range also support
the above description.\cite{Araujo} These energy barriers increase as the
temperature is lowered, and therefore the capability of the measuring field
to induce the structural transition from non-FM to a FM state diminishes.
This is responsible for the lower magnetization in the {\it M}(1-0) data of
Fig. 5c. However, the same magnetic field applied in the FCC mode can
locally prevent the formation of the secondary phase, giving rise to an
enhanced amount of the FM phase with respect to that obtained in the {\it M}(%
{\it x-y}) mode.

\begin{center}
\begin{figure}[tbp]
\hspace{-0.0cm}
\epsfxsize=8.50cm
\centerline{\epsffile{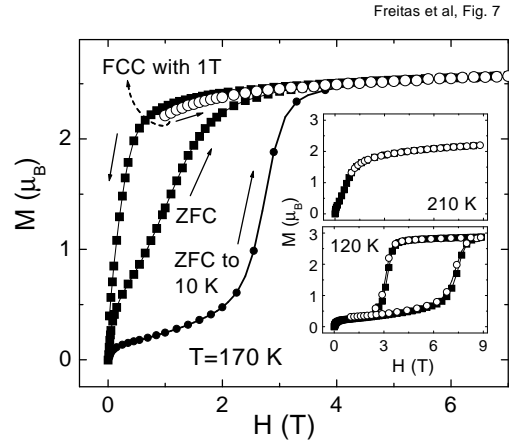}}
\caption{Magnetization as a function of field of La$_{0.5}$Ca$_{0.5}$MnO$%
_{3}$, sample E. The main panel shows results measured after
zero-field-cooling (ZFC) to T = 170 K (closed squares), after ZFC to T = 10
K and subsequently warmed to 170 K (closed circles), and after
field-cooled-cooling (FCC) with H = 1 T to 170 K (open circles). The insets
show data measured after ZFC (closed squares) and FCC (open circles) to T =
210 and 120 K.
\label{Fig7} 
}
\end{figure}
\end{center}
The magnetic properties of the intermediate phase separated state become
evident through measurements of {\it M} vs. {\it H} at T = 170 K (main panel
of Fig. 7). The data recorded after FCC with {\it H} = 1 T starts from a
nearly saturated magnetization value, reflecting that the cooling field
almost totally inhibits the formation of the secondary phase. Instead, the
results obtained after ZFC to 170 K show a low initial magnetic component,
and a smooth metamagnetic-like transition, completed at {\it H }$\sim $ 3-4
T. An additional curve is displayed, in which the sample is cooled to 10 K,
and subsequently warmed to the target temperature. In this case the majority
phase is the low temperature AFM-CO phase, which can be transformed to a FM
state in a sharp metamagnetic transition, as observed in the figure.
Finally, below T$_{CO}$ most of the material falls into the CO-AFM state,
which is robust with respect to the application of moderated magnetic
fields, and the relative phase fractions are not longer controlled by the
measuring field but only by the cooling field. At 120 K, both ZFC and FCC 
{\it M}({\it H}) curves (bottom inset of Fig.7) display the same overall
behavior, with an abrupt field-induced CO-AFM to FM metamagnetic transition
at {\it H} $\simeq $ 7.4 T The data obtained after ZFC and FCC conditions
differ only in the magnetization values just above 1 T, due to the FFE
effect of the cooling field.

\section{Conclusions}

The present investigation addresses the effects of an applied magnetic field
in the magnetic phases of La$_{0.5}$Ca$_{0.5}$MnO$_{3}$, one of the
prototype compounds for studying PS effects in manganites. The close
interplay between the FM fraction and the applied magnetic field is crucial
for interpreting the magnetization of this phase separated system. We have
used a particular experimental protocol related to the application of the
magnetic field, which reveal the phase separated nature of the compound, due
to the distinguishable effects of the cooling and measuring magnetic fields.
The presented magnetization results show the existence of three well
differentiated PS regimes: `soft PS' for T$_{C}$ 
\mbox{$>$}%
T 
\mbox{$>$}%
T$_{o}$, `intermediate' PS for T$_{o}$ 
\mbox{$>$}%
T 
\mbox{$>$}%
T$_{CO}$, and `hard PS' for T 
\mbox{$<$}%
T$_{CO}$. These regimes were identified following the system's response to
an applied magnetic field, and are consistent with neutron scattering
results.\cite{Huang}

The actual existence of a phase separated state close below T$_{C}$ is not
directly revealed by our measurements, but was addressed by other authors.%
\cite{Mori,Chen-Cheong,Rivadulla,AllodiPRL} In this region the magnetic
response of the system is the same in a wide range of magnetic fields,
irrespective of the way in which the field is applied. This seems to
indicate that the coexisting phases differ in their magnetic properties
(paramagnetic-like and FM) but do not display energy barriers at the phase
boundaries. This fact characterizes the soft PS regime, in which the FM
phase can grow almost freely with the application of a magnetic field,
whichever mode is employed. The intermediate region has its onset at T$_{o}$%
, where magnetization curves obtained in the FCC and {\it M}({\it x-y})
modes begin to separate. The different response of the system to the cooling
and the measuring fields is consistent with the fact that the phase
separated state in this temperature range is of a structural nature,\cite
{Huang} with some degree of charge and/or orbital ordering in the secondary
phase.\cite{Mahendiran,Moritomo} A moderate applied magnetic field ($\sim $
1 T) is successful in preventing the formation of the secondary phase in the
FCC mode. On the other hand, in the {\it M}({\it x-y}) mode the magnetic
field has to induce a structural transition to accomplish the growth of the
FM phase against the non-FM one, overcoming the energy barriers at the phase
interfaces, therefore much less FFE is achieved. It is worth noting that the
preceding statements are only valid for measuring fields within a definite
window, outside which magnetic measurements are unable to reveal the phase
separated nature of the system at this intermediate temperature range. With
low magnetic field values ({\it H} $\lesssim $ 0.1 T) even the cooling field
does not induce FFE, while at relatively high values ({\it H} $\gtrsim $ 3
T) the measuring field alone promotes a complete metamagnetic transition,
transforming the system in to a nearly homogeneous FM state.

The hard PS regime which appears below T$_{CO}$ is the most studied in
previous papers. It consist of minority FM regions embedded in an CO-AFM
host with CE-type structure, a more robust state with respect to the
application of a magnetic field. However, relatively low cooling fields can
produce small variations of the FM fraction, which in turn produces huge
changes in the resistivity when the FM fraction is near the percolation
threshold. The transport data presented provide clear evidence for the
percolative nature of the low field low-temperature colossal
magnetoresistance. At temperatures well below T$_{CO}$, an applied cooling
field is solely responsible for a sizeable decrease in electrical transport.
On the other hand, there is no appreciable magnetoresistance in measurements
with the {\it R}({\it x}-0) mode, since the measuring field can not produce
FFE in the hard PS region. Both transport and magnetization data also show
the drastic effect of the cooling field in lowering the CO transition
temperature, whereas no reduction in T$_{CO}$ is observed in the {\it M}(%
{\it x}-0) measurements with fields up to $\sim $ 3 T.

Summarizing, the measured magnetization values in phase separated manganites
depend strongly on the magnetic field applied during the measurements, and
on the field applied while cooling the sample. We were able to determine
through magnetization data the onset at T$_{o}$ of the intermediate PS
region, and established the existence of a critical cooling field above
which FFE occurs. The overall results show how {\it turn-on-turn-off} -type
magnetization measurements can be employed to reveal the phase separated
nature of manganite compounds. Needless to say, more powerful tools such as
microscopic techniques are required to fully characterize the coexisting
phases at the various temperature ranges.

\section{Acknowledgments}

We thank G. Polla and G. Leyva for sample preparation. This work was
partially financed by PRONEX/FINEP/CNPq (contract no 41.96.0907.00) and
CONICET PEI 125/98. Additional support was given by FAPERJ, and by
Antorchas, Balseiro, and Vitae Foundations.

* Corresponding author. Electronic address: luisghiv@if.ufrj.br


\begin{references}
\bibitem{Salamon}  M.B. Salamon and M. Jaime, Rev. Mod. Phys. {\bf 73}, 583
(2001).

\bibitem{Moreo}  A. Moreo, S. Yunoki, and E. Dagotto, Science {\bf 283},
2034 (1999).

\bibitem{Uehara}  M. Uehara, S. Mori, C.H. Chen, S.W. Cheong Nature (London) 
{\bf 399}, 560 (1999).

\bibitem{Dagotto}  E. Dagotto, T. Hotta, and A. Moreo, Physics Reports -
Rev. Section of Physics Letters {\bf 344}, 1 (2001).

\bibitem{Schiffer}  P. Schiffer et al, Phys. Rev. Lett. {\bf 75}, 3336
(1995).

\bibitem{Huang}  Q. Huang, J.W. Lynn, R.W. Erwin, A. Santoro, D.C. Dender,
V.N. Smolyaninova, K. Ghosh, and R.L. Greene, Phys. Rev B {\bf 61}, 8895
(2000).

\bibitem{Levy}  P. Levy, F. Parisi, G. Polla, D. Vega, G. Leyva, H. Lanza,
R.S. Freitas and L. Ghivelder, Phys. Rev. B {\bf 62}, 6437 (2000).

\bibitem{Mori}  S. Mori, C.H. Chen, and S-W. Cheong, Phys. Rev. Lett. {\bf 81%
}, 3972 (1998).

\bibitem{Allodi}  G. Allodi, R. De Renzi, M. Solzi, K. Kamenev, G.
Balakrishnan and M. W. Pieper, Phys. Rev. B {\bf 61}, 5924 (2000).

\bibitem{Yoshinari}  Y. Yoshinari, P. C. Hammel, J. D. Thompson, and S-W.
Cheong, Phys. Rev. B {\bf 60}, 9275 (1999).

\bibitem{Parisi}  F. Parisi, P. Levy, L. Ghivelder, G. Polla, and D. Vega,
Phys. Rev. B {\bf 63}, 144419 (2001).

\bibitem{Chen-Cheong}  C. H. Chen and S-W. Cheong, Phys. Rev. Lett. {\bf 76}%
, 4042 (1996).

\bibitem{Simopoulos}  A. Simopoulos, G. Kallias, E. Devlin, and M. Pissas,
Phys.Rev B {\bf 63}, 054403 (2001).

\bibitem{Rivadulla}  F. Rivadulla, M. Freita-Alvite, M. A. Lopez-Quintela,
L. E. Hueso, D. R. Miguens, P. Sande, J. Rivas, cond-mat/0105088
(unpublished).

\bibitem{Mahendiran}  R. Mahendiran, A. Maignan, C. Martin, M. Hervieu, B.
Raveau, cond-mat/0107448 (unpublished).

\bibitem{Roy}  M. Roy, J. F. Mitchell, A. P. Ramirez and P. Schiffer, Phys.
Rev. B {\bf 58}, 5185 (1998).

\bibitem{Ritter}  C. Ritter, R. Mahendiran, M. R. Ibarra, L. Morellon, A.
Maignan, B. Raveau and C. N. R. Rao, Phys. Rev. B {\bf 61}, R9229 (2000).

\bibitem{DagottoPRL}  M. Mayr, A. Moreo, J. A. Verges, J. Arispe, A.
Feiguin, and E. Dagotto, Phys. Rev. Lett. {\bf 86}, 135 (2001).

\bibitem{Sacanell}  J. Sacanell, P. Levy, L. Ghivelder, G. Polla, D. Vega
and F. Parisi, Physica B (in press).

\bibitem{Younghun}  Younghun Jo, J.-G. Park, Chang Seop Hong, N.H. Hur, and
H.C. Ri, Phys. Rev. B {\bf 63}, 172413 (2001).

\bibitem{Goodenough}  E. O. Wollan and W. C. Koehler, Phys. Rep. {\bf 100},
545 (1955); J. B. Goodenough, ibid {\bf 100}, 564 (1955).

\bibitem{Uehara-eulett}  M. Uehara and S-W. Cheong. Europhys. Lett. {\bf 52}%
, 674 (2000).

\bibitem{Mahendiran-SSC}  R. Mahendiran, M.R. Ibarra, A. Maignan, C. Martin,
B. Raveau and A. Hernando, Solid State Commun. {\bf 111}, 525 (1999).

\bibitem{Araujo}  J. Lopez, P. N. Lisboa-Filho, W. A. C. Passos, W. A.
Ortiz, F. M. Araujo-Moreira , O. F. de Lima, D. Schaniel, and K. Ghosh,
Phys. Rev. B {\bf 63}, 224422 (2001).

\bibitem{AllodiPRL}  G. Allodi, R. De Renzi, F. Licci and M. W. Pieper,
Phys. Rev. Lett. {\bf 81}, 4736 (1998).

\bibitem{Moritomo}  Y. Moritomo, Phys. Rev. B {\bf 60}, 10374 (1999).
\end{references}
\end{document}